\documentclass[11pt,a4paper]{article}
\usepackage{jheppub}

\usepackage{epstopdf}

\newcommand{\dd}{{\text{d}}}

\title{\boldmath
Matter Effects of Sterile Neutrino in Light of Renormalization-Group Equations}

\author{Shuge Zeng,} 
\author{Fanrong Xu \footnote{Corresponding author.}}

\affiliation
{Department of Physics and Siyuan Laboratory, Jinan University,\\
$\;$Guangzhou 510632, P.R. China}

\emailAdd{sgzeng@stu2021.jnu.edu.cn}
\emailAdd{fanrongxu@jnu.edu.cn}

\abstract{

The renormalization-group equation (RGE) approach to neutrino matter effects  is further developed in this work. 
We derive a complete set of differential equations for effective mixing elements, masses and Jarlskog-like invariants
  in presence of a light sterile neutrino. 
The evolutions of  mixing elements as well as Jarlskog-like invariants are obtained
by numerically solving these differential equations. 
We calculate terrestrial matter effects in long-baseline (LBL) experiments, taking NOvA, T2K and DUNE as examples.
 In both three-flavor and four-flavor frameworks,
electron-neutrino survival probabilities  as well as the day-night asymmetry of solar neutrino 
are also evaluated as a further examination of the RGE approach.
 }


\begin{document}
\maketitle
\flushbottom


\newpage

\section{Introduction}
\label{sec:Intro}

Now it is well established that active neutrinos are massive with tiny masses and oscillate among
different flavor states.  The measurement of oscillation parameters are improved noticeabely in the last 
two decades.  On the other hand, as  hypothetical particles, sterile neutrinos do not couple directly to the gauge bosons
and can only participate in weak interactions through mixing with active neutrinos. Theoretically their mass and number are 
unconstrained, and have been quested in experiments for a long time.
The existence of eV mass scale sterile neutrinos has been motivated by several low energy anomalies 
which cannot be accounted for by the standard three-flavor active neutrino framework. 
These involve short-baseline (SBL) experiments with neutrinos from accelerators, reactors and radioactive sources.

Two most well-known accelerator-based experiments are LSND and MiniBooNE.  
The LSND anomaly is the $3.8\sigma$ excess of events compatible to $\bar{\nu}_e$ appearance in a $\bar{\nu}_\mu$ beam observed by the LSND experiment \cite{LSND:2001aii},  while the MiniBooNE anomaly is the $4.8\sigma$ excess of electron-like events in the MiniBooNE experiment observed in both $\nu_\mu$ and $\bar{\nu}_\mu$ beams \cite{MiniBooNE:2018esg}.  
As a direct test of LSND anomaly, the JSNS$^2$ experiment has started data taking \cite{Hino:2021uwz} at J-PARC in Japan.
With the same neutrino beam as MiniBooNE but superior event reconstruction capabilities, 
 the MicroBooNE collaboration recently has released results scrutinizing the MiniBooNE anomaly, known as the MiniBooNE low-energy excess (MBLEE). 
 A first MicroBooNE analysis disfavors that the MBLEE is due to underestimated production of $\Delta$ baryons followed by decays to photons at a significance of $94.8\%$ C.L. \cite{MicroBooNE:2021zai} and concluded that 
 ``no excess of $\nu_e$ 
 events is observed'' \cite{MicroBooNE:2021rmx}. However, a recent work shows quantitatively that 
 the MicroBooNE collaboration
``unquestionably do not probe the full parameter space of sterile neutrino models hinted at by MiniBooNE, nor do they probe the 
$\nu_e$ interpretation of the MiniBooNE excess in a model-independent way'' \cite{Arguelles:2021meu}, which indicates that it is still early 
to exclude eV scale sterile neutrinos.
The reactor antineutrino anomaly (RAA), is about the $ 6\%$ deficit (a $3\sigma$ effect) of $\bar{\nu}_e$
 in reactor VSBL (very short-baseline, $< 100$ m) experiments resulting from the re-evaluation of the reactor antineutrino flux \cite{Huber:2011wv, Mention:2011rk}, compared to the new prediction in 2011 \cite{Mueller:2011nm}.
Many VSBL experiments have been created and  progresses have been made, 
including NEOS\cite{NEOS:2016wee}, DANSS\cite{Danilov:2019aef}, STEREO\cite{STEREO:2020hup}, 
Neutrino-4\cite{Serebrov:2020kmd}, PROSPECT\cite{PROSPECT:2020sxr} and so on. 
The gallium anomaly occurred in the two experiments
 GALLEX \cite{Hampel:1997fc, Kaether:2010ag} and SAGE \cite{Abdurashitov:1998ne,Abdurashitov:2005tb}, where a deficit of $16\%$  events between
observations and theoretical calculation \cite{Schwetz:2013naa} was found. Though the latest estimation carried out in 2019 using new nuclear shell-model wave functions diminish the significance from $3 \sigma$ to $2.3\sigma$ \cite{Kostensalo:2019vmv}, 
recently BEST \cite{Barinov:2021asz} reaffirmed the gallium anomaly with a larger deficit $20\!-\!24\%$. 
More can be referred to the recent review papers \cite{Dasgupta:2021ies, Athar:2021xsd}.

Regardless of the sources, baselines and energy ranges of these experiments, all of the results involving 
above experiments can be understood individually via SBL neutrino oscillations driven by 
$\Delta m^2\sim 1 \textrm{eV}^2$,  requiring a fourth neutrino mass eigenstate to account for this higher mass-squared difference.
Hence  the existence of eV mass scale light sterile neutrinos  is still an open question to be answered
at this stage.

The Mikheyev-Smirnov-Wolfenstein (MSW) matter effects \cite{Wolfenstein:1977ue,Mikheev:1986wj} plays an essential role
in the determination of neutrino mass ordering and leptonic CP-violating phase  via the long-baseline (LBL) accelerator neutrino
experiments. When the neutrino beam propagates in the Earth matter for a long distance, the MSW effect becomes 
crucially important.  
In presence of a light sterile neutrino, more degrees of freedom enter the MSW effect even in the propagation of 
active flavor neutrinos. For the calculation of matter effects involving a light sterile neutrino,  
different methods \cite{Klop:2014ima, Li:2018ezt, Yue:2019qat}
 have been developed in recent years.
 
The idea of renormalization-group equations (RGEs) has been widely applied in  
quantum fields theories\cite{StueckelbergdeBreidenbach:1952pwl, Gell-Mann:1954yli}, 
solid-state physics\cite{Wilson:1971bg, Wilson:1971dh}, and other fields of modern physics.
There have been 
 attempts to make a connection between RGE and neutrino matter effects in the past few years\cite{Chiu:2010da, Chiu:2017ckv}. 
Recently, more progress has been made \cite{Xing:2018lob, Zhou:2020iei, Wang:2019yfp}.
By taking the matter term $a$ (see Section \ref{sec:ana.}) as the role of renormalization energy scale, 
a close set of  differential equations 
 describing
the evolution of effective mixing matrix and effective mass-square differences in three-flavor case has been
established \cite{Xing:2018lob}.
The analytical solutions for the mixing and masses  are further given in \cite{Wang:2019yfp}.
One may naturally wonder whether it is accessible to extend the RGE framework to include the sterile neutrino contribution.
In addition to developing the form theoretically, 
it is also interesting to realize the
 practical application of this new method in
neutrino phenomenology.
These topics are supposed to be investigated in current work.
Although it does not participate weak interactions directly, the sterile neutrino contribution in Hamiltonian is brought in through 
subtracting a global identical term produced via charged lepton neutral-current, denoted as
 mass term $a'$ (see Section \ref{sec:ana.}). In the scenario of constant density matter, two mass terms of electron-type  and
 sterile neutrino is connected by a constant $k$.  
Then we can continue working in the framework of RGE approach \cite{Xing:2018lob}, and  extend the evolution of mixing matrix
and mass-square differences with respect to mass term $a$ (or equivalently neutrino beam energy $E$), 
including the sterile neutrino contribution.
After analytically deriving the completed set of differential equations,  we further explore their 
solutions numerically. In principle, oscillation probabilities should be calculable in the RGE approach, 
which now is realized
 here for the 
first time  in both cases
with purely three active neutrinos as well as with one more sterile neutrino.
We take three LBL accelerator neutrino experiments, NOvA, T2K and DUNE, as typic examples to test the validity of RGE 
approach in practical numerical analysis and find that the results perfectly agree with the analytical approach \cite{Li:2018ezt}. 
In the adiabatic approximation of solar matter and slab approximation of Earth matter, we also calculate 
the day-night asymmetry observed in the surface of Earth, as a further application of RGE method.

The remaining part of this paper is organized as follows. In Section \ref{sec:ana.}, we present the derivation details and
exact forms 
of differential equations of 
masses and mixing matrix, 
as well as oscillation probabilities.
After exhibiting the explicit evolution of mixing matrix elements involving a light sterile neutrino in first part of Section \ref{sec:num},
we further apply the methodology to Earth matter effects in three LBL experiments and  solar matter effects in name of day-night asymmetry. 
We conclude the validity of RGE approach in Section \ref{sec:con} and related input parameters are summarized in Appendix \ref{app:input}.


\section{Analytical formulae with RGE}
\label{sec:ana.}

It is known that physical observables are independent of energy scale while parameters 
describing the observables depend on scale themselves, and 
renormalization-group equations  (RGEs) depict the evolutions of these parameters with respect to the energy scale. 
With the general idea of RGE, a new method is developed to describe 
flavor mixing parameters evolve when neutrinos propagate in a
medium in the framework of three-flavor neutrino oscillation\cite{Xing:2018lob}.
As introduced in Sec.\ref{sec:Intro}, the existence of eV-scale light sterile neutrinos is still 
an open question and the calculation of matter effects with sterile neutrino is meaningful,
especially in the measurement of mass ordering and leptonic CP-violating phase(es).
In this section, we will make an extension of the RGE approach to contain the case in presence of a light sterile neutrino in addition to three active flavors. 

\subsection{Master equations for effective masses and mixing matrix}
Similar to the standard three-flavor case, 
the effective
Hamiltonian with sterile neutrino contribution is given as
\begin{equation}
H_m = \frac{1}{2E} \left[ U \left(\begin{array}{cccc}
m_1^2 & 0 & 0 & 0\\
0 & m_2^2 & 0 & 0 \\
0 & 0 & m_3^2 & 0\\
0 & 0 & 0 & m_4^2
\end{array}\right) U^\dagger +
\left(\begin{array}{cccc}
a & 0 & 0 & 0 \\
0 & 0 & 0 & 0\\
0 & 0 & 0 & 0\\
0 & 0 & 0 & a'
\end{array}\right)\right] \equiv \frac{1}{2E} V
\left(\begin{array}{cccc}
\tilde{m}_1^2 & 0 & 0 &0\\
0 & \tilde{m}_2^2 & 0 &0\\
0 & 0 & \tilde{m}_3^2 &0\\
0 & 0 & 0& \tilde{m}^2_4\\
\end{array}\right) V^\dagger,
\label{eq:Hamiltonian}
\end{equation}
where $a\equiv 2\sqrt{2} G_F N_e E$ with $N_e$ being the net electron number density and $E$
the neutrino beam energy, mass terms $m_i^2$ and $\tilde{m}_i^2$ in vacuum and matter, 
respectively.
For the mixing matrices, the mixing in vacuum (PMNS matrix) is denoted as $U$ \footnote{
The convention and inputs to parameterize PMNS matrix $U$ 
is introduced in Appendix \ref{app:input}. } 
and
$V$ stands for the effective mixing incorporating matter effects.
Though it does not take part in weak interaction itself, the sterile neutrino has an
 matter effects term $a'$ here due to the subtraction of a global identical term
 produced via neutral-current interaction of active neutrinos. In general, $a'$ dependents 
 on the medium neutrino beam propagating. In the scenario of constant density matter, 
 we can further write down $a'=ka$ and the medium information is encoded in the matter parameter $k$.

We take  $a$ as the role of energy scale in RGE 
similar as \cite{Xing:2018lob}.
Differentiating both sides of Eq.(\ref{eq:Hamiltonian})
with respect to $a$,  we have
\begin{eqnarray}
\dot{D} + \left[ V^\dagger \dot{V}, D\right]
&=&V^\dagger \left(\begin{array}{cccc}
1 & 0 & 0& 0\\
0 & 0 & 0& 0\\
0 & 0 & 0& 0\\
0 & 0 & 0& k
 \end{array} \right) V\label{eq:diag}\\
 &=&
 {\scriptsize{
\left(\begin{array}{cccc}
|V_{e1}|^2 + k|V_{s1}|^2 & V_{e1}^* V_{e2}+ k V_{s1}^* V_{s2} & V_{e1}^* V_{e3}+ k V_{s1}^* V_{s3} & V_{e1}^*V_{e4}+ k V_{s1}^* V_{s4}\\
V_{e2}^* V_{e1} + k V_{s2}^* V_{s1}& |V_{e2}|^2+ k|V_{s2}|^2 & V_{e2}^* V_{e3}+ k V_{s2}^* V_{s3}&
 V_{e2}^*V_{e4}+ k V_{s2}^* V_{s4}\\
V_{e3}^* V_{e1} + k V_{s3}^* V_{s1}& V_{e3}^* V_{e2}+ k V_{s3}^* V_{s2} &|V_{e3}|^2+ k|V_{s3}|^2 &
 V_{e3}^*V_{e4}+ k V_{s3}^* V_{s4}\\
V_{e4}^* V_{e1} + k V_{s4}^* V_{s1}&V_{e4}^* V_{e2}+ k V_{s4}^* V_{s2} & V_{e4}^* V_{e3}+ k V_{s4}^* V_{s3}&
 |V_{e4}|^2+ k|V_{s4}|^2
 \end{array}\right),
}}\nonumber
\end{eqnarray}
where $D\equiv {\textrm{diag}} \left\{ \tilde{m}_1^2, \tilde{m}_2^2, \tilde{m}_3^2, \tilde{m}_4^2\right\}$.
Apparentely, 
the first term in the left-handed side of  Eq. (\ref{eq:diag}) is diagonal 
while diagonal entries of the second term vanish. 
Extracting the diagonal entries in both sides, we can firstly get
a relation
\begin{equation*}
\frac{ \dd\tilde{m}_i^2}{\dd a} = |V_{ei}|^2 + k |V_{si}|^2,
\end{equation*}
which further leads to the first order differential equations 
\begin{equation}
\frac{\dd}{\dd a}\Delta\tilde{m}^2_{ij} = (|V_{ei}|^2 - |V_{ej}|^2)
+ k(|V_{si}|^2 - |V_{sj}|^2)
\label{eq:Mass-diff}
\end{equation}
with the effective mass square difference  $\Delta\tilde{m}^2_{ij}\equiv \tilde{m}_i^2 -\tilde{m}_j^2$. 
Similarly, by comparing the off-diagonal entries in Eq.  (\ref{eq:diag}), one has
\begin{equation}
\sum_\alpha V_{\alpha i}^* \dot{V}_{\alpha j} = (\Delta\tilde{m}^2_{ji})^{-1}
\left( V_{ei}^* V_{ej} + k V_{si}^* V_{sj}\right).
\label{eq:offD}
\end{equation}
Multiply both sides of the orthogonal relation
\begin{equation*}
\sum_{j\neq i} V_{\alpha j}^* V_{\beta j} = \delta_{\alpha \beta} - V_{\alpha i}^* V_{\beta i},
\end{equation*}
 with $\dot{V}_{\alpha i}$, and take a summation over $\alpha$ (so that 
 $\sum\limits_{\alpha} \dot{V}_{\alpha i} V_{\alpha i}^* = 0$
 ), we can write down the
 differential equation for effective mixing matrix as
\begin{equation}
\dot{V}_{\beta i}=
\sum_{j\neq i} V_{\beta j} (V_{ej}^* V_{ei} + k V_{sj}^* V_{si})(\Delta\tilde{m}^2_{ij})^{-1},
\label{eq:Vdot}
\end{equation}
in which Eq. (\ref{eq:offD}) is applied and summation is not taken for index $i$. 
Due to the fact
\begin{equation*}
\sum_\beta \dot{V}^*_{\beta i} V_{\beta i} V^*_{\alpha i}V_{\alpha i}
+\sum_\beta V^*_{\alpha i} V_{\alpha i} \dot{V}_{\beta i} V^*_{\beta i}
=|V_{\alpha i}|^2 \sum_\beta (\dot{V}^*_{\beta i} V_{\beta i} + \dot{V}_{\beta i} V^*_{\beta i})=0,
\end{equation*}
the differential equation of mixing matrix can also be written  
 in terms of the square of mixing matrix,
\begin{equation}
\frac{\dd }{\dd a}|V_{\alpha i}|^2 
= \left(\frac{\dd }{\dd a} V_{\alpha i}^*\right) V_{\alpha i}+ V^*_{\alpha i} \frac{\dd }{\dd a}V_{\alpha i}
= 2\sum_{j\neq i} {\text{Re}}\left[ V_{\alpha j}V^*_{\alpha i}
(V^*_{ej} V_{ei} + k V^*_{sj} V_{si})\right] (\Delta\tilde{m}^2_{ij})^{-1},
\label{eq:Vsquare}
\end{equation}
this compact form is phase-independent and brings convenience in some parts of
 following calculation. 

\subsection{The detailed differential equations}
\label{subsec:eq}
The master equations (\ref{eq:Mass-diff})
 and (\ref{eq:Vsquare}) (or  (\ref{eq:Vdot})) form a closed and complete equation group,
 which contains 19 independent differential equations. 
 To have a global impression, it is helpful to exhibit some necessary differential equations explicitly. 
Three  independent equations on mass square differences  
can be  detailed from 
equation (\ref{eq:Mass-diff}),
\begin{equation}
\begin{split}
& \frac{\dd}{\dd a}\Delta \tilde{m}_{12}^2 = (|V_{e1}|^2-|V_{e2}|^2) + k ( |V_{s1}|^2-|V_{s2}|^2), \\
& \frac{\dd}{\dd a}\Delta \tilde{m}_{13}^2 = (|V_{e1}|^2-|V_{e3}|^2) + k ( |V_{s1}|^2-|V_{s3}|^2),   \\
& \frac{\dd}{\dd a}\Delta \tilde{m}_{14}^2 = (|V_{e1}|^2-|V_{e4}|^2) + k ( |V_{s1}|^2-|V_{s4}|^2).
\end{split}
\label{eq:EffMass}
\end{equation}
The right handed side of equation (\ref{eq:EffMass}) indicates that  the evolution property of mass square difference relies on square of mixing matrix entries together with the matter parameter $k$. As part of the closed equation group, 
the detailed differential equations for $\alpha=e$ can be read from equation (\ref{eq:Vsquare}),
\begin{equation}
\begin{split}
& \frac{\dd}{\dd a}|V_{e1}|^2 = 2 |V_{e1}|^2 \left[ |V_{e2}|^2 (\Delta \tilde{m}_{12}^2)^{-1}
-|V_{e3}|^2 (\Delta \tilde{m}_{31}^2)^{-1}-|V_{e4}|^2 (\Delta \tilde{m}_{41}^2)^{-1}
\right]\\
&\hspace{1.5cm} +2 k {\text{Re}} \left[ V_{e1}^* V_{s1}\Big( V_{e2}V_{s2}^* (\Delta \tilde{m}_{12}^2)^{-1}
-V_{e3}V_{s3}^* (\Delta \tilde{m}_{31}^2)^{-1}-V_{e4}V_{s4}^* (\Delta \tilde{m}_{41}^2)^{-1}\Big)\right],\\
& \frac{\dd}{\dd a}|V_{e2}|^2 = 2 |V_{e2}|^2 \left[ |V_{e3}|^2 (\Delta \tilde{m}_{23}^2)^{-1}
-|V_{e4}|^2 (\Delta \tilde{m}_{42}^2)^{-1}-|V_{e1}|^2 (\Delta \tilde{m}_{12}^2)^{-1}
\right]\\
&\hspace{1.5cm} + 2k {\text{Re}} \left[ V_{e2}^* V_{s2}\Big( V_{e1}V_{s1}^* (\Delta \tilde{m}_{21}^2)^{-1}
-V_{e3}V_{s3}^* (\Delta \tilde{m}_{32}^2)^{-1}-V_{e4}V_{s4}^* (\Delta \tilde{m}_{42}^2)^{-1}\Big)\right],\\
& \frac{\dd}{\dd a}|V_{e3}|^2 = 2 |V_{e3}|^2 \left[ |V_{e4}|^2 (\Delta \tilde{m}_{34}^2)^{-1}
-|V_{e1}|^2 (\Delta \tilde{m}_{13}^2)^{-1}-|V_{e2}|^2 (\Delta \tilde{m}_{23}^2)^{-1}
\right]\\
&\hspace{1.5cm} + 2k {\text{Re}} \left[ V_{e3}^* V_{s3}\Big( V_{e4}V_{s4}^* (\Delta \tilde{m}_{34}^2)^{-1}
-V_{e1}V_{s1}^* (\Delta \tilde{m}_{13}^2)^{-1}-V_{e2}V_{s2}^* (\Delta \tilde{m}_{23}^2)^{-1}\Big)\right],\\
& \frac{\dd}{\dd a}|V_{e4}|^2 = 2 |V_{e4}|^2 \left[ |V_{e1}|^2 (\Delta \tilde{m}_{41}^2)^{-1}
-|V_{e2}|^2 (\Delta \tilde{m}_{24}^2)^{-1}-|V_{e3}|^2 (\Delta \tilde{m}_{34}^2)^{-1}
\right]\\
&\hspace{1.5cm} + 2k {\text{Re}} \left[ V_{e4}^* V_{s4}\Big( V_{e1}V_{s1}^* (\Delta \tilde{m}_{41}^2)^{-1}
-V_{e2}V_{s2}^* (\Delta \tilde{m}_{24}^2)^{-1}-V_{e3}V_{s3}^* (\Delta \tilde{m}_{34}^2)^{-1}\Big)\right].
\end{split}\label{eq:VeSquare}
\end{equation}
The evolution equations of $|V_{\alpha i}|^2$, taking $|V_{e i}|^2$ as an example, contain more information than
 the ones of $\Delta \tilde{m}^2_{ij}$ for 
their entanglement with mass-square differences, as well as  individual mixing matrix elements.
From the known the general expression for the evolution of $V_{\alpha i}$ with respect to $a$ equation (\ref{eq:Vdot})), we 
can further write down explicitly those  related to
electron-type neutrino and sterile neutrino, giving
\begin{equation}
\begin{split}
\frac{\dd V_{e1}}{\dd a}=\,&(V_{e1}|V_{e2}|^{2}+kV_{e2}V_{s1}V_{s2}^{\ast})(\Delta\tilde{m}_{12}^2)^{-1}
                  -(V_{e1}|V_{e3}|^{2}+kV_{e3}V_{s1}V_{s3}^{\ast})(\Delta\tilde{m}_{31}^2)^{-1}\\
                  &-(V_{e1}|V_{e4}|^{2}+kV_{e4}V_{s1}V_{s4}^{\ast})(\Delta\tilde{m}_{41}^2)^{-1},\\
\frac{\dd V_{e2}}{\dd a}=\,&(V_{e2}|V_{e1}|^{2}+kV_{e1}V_{s2}V_{s1}^{\ast})(\Delta\tilde{m}_{21}^2)^{-1}
                  -(V_{e2}|V_{e3}|^{2}+kV_{e3}V_{s2}V_{s3}^{\ast})(\Delta\tilde{m}_{32}^2)^{-1}\\
                  &-(V_{e2}|V_{e4}|^{2}+kV_{e4}V_{s2}V_{s4}^{\ast})(\Delta\tilde{m}_{42}^2)^{-1},\\
\frac{\dd V_{e3}}{\dd a}=\,&(V_{e3}|V_{e1}|^{2}+kV_{e1}V_{s3}V_{s1}^{\ast})(\Delta\tilde{m}_{31}^2)^{-1}
                  -(V_{e3}|V_{e2}|^{2}+kV_{e2}V_{s3}V_{s2}^{\ast})(\Delta\tilde{m}_{23}^2)^{-1}\\
                  &-(V_{e3}|V_{e4}|^{2}+kV_{e4}V_{s3}V_{s4}^{\ast})(\Delta\tilde{m}_{43}^2)^{-1},\\
\frac{\dd V_{e4}}{\dd a}=\,&(V_{e4}|V_{e1}|^{2}+kV_{e1}V_{s4}V_{s1}^{\ast})(\Delta\tilde{m}_{41}^2)^{-1}
                  +(V_{e4}|V_{e2}|^{2}+kV_{e2}V_{s4}V_{s2}^{\ast})(\Delta\tilde{m}_{42}^2)^{-1}\\
                  &+(V_{e4}|V_{e3}|^{2}+kV_{e3}V_{s4}V_{s3}^{\ast})(\Delta\tilde{m}_{43}^2)^{-1},
\end{split}\label{eq:Vei}
\end{equation}
and 
\begin{equation}
\begin{split}
\frac{\dd V_{s1}}{\dd a}=\,&(V_{s2}V_{e2}^{\ast}V_{e1}+k|V_{s2}|^{2}V_{s1})(\Delta\tilde{m}_{12}^2)^{-1}
                  -(V_{s3}V_{e3}^{\ast}V_{e1}+k|V_{s3}|^{2}V_{s1})(\Delta\tilde{m}_{31}^2)^{-1}\\
                  &-(V_{s4}V_{e4}^{\ast}V_{e1}+k|V_{s4}|^{2}V_{s1})(\Delta\tilde{m}_{41}^2)^{-1},\\
\frac{\dd V_{s2}}{\dd a}=\,&(V_{s1}V_{e1}^{\ast}V_{e2}+k|V_{s1}|^{2}V_{s2})(\Delta\tilde{m}_{21}^2)^{-1}
                  -(V_{s3}V_{e3}^{\ast}V_{e2}+k|V_{s3}|^{2}V_{s2})(\Delta\tilde{m}_{32}^2)^{-1}\\
                  &-(V_{s4}V_{e4}^{\ast}V_{e2}+k|V_{s4}|^{2}V_{s2})(\Delta\tilde{m}_{42}^2)^{-1},\\
\frac{\dd V_{s3}}{\dd a}=\,&(V_{s1}V_{e1}^{\ast}V_{e3}+k|V_{s1}|^{2}V_{s3})(\Delta\tilde{m}_{31}^2)^{-1}
                  -(V_{s2}V_{e2}^{\ast}V_{e3}+k|V_{s2}|^{2}V_{s3})(\Delta\tilde{m}_{23}^2)^{-1}\\
                  &-(V_{s4}V_{e4}^{\ast}V_{e3}+k|V_{s4}|^{2}V_{s3})(\Delta\tilde{m}_{43}^2)^{-1},\\
\frac{\dd V_{s4}}{\dd a}=\,&(V_{s1}V_{e1}^{\ast}V_{e4}+k|V_{s1}|^{2}V_{s4})(\Delta\tilde{m}_{41}^2)^{-1}
                  +(V_{s2}V_{e2}^{\ast}V_{e4}+k|V_{s2}|^{2}V_{s4})(\Delta\tilde{m}_{42}^2)^{-1}\\
                  &-(V_{s3}V_{e3}^{\ast}V_{e4}+k|V_{s3}|^{2}V_{s4})(\Delta\tilde{m}_{43}^2)^{-1}.\\
\end{split}
\end{equation}
Although equations   (\ref{eq:Vei})  and (\ref{eq:VeSquare}) 
are not independent ones, here for 
 the purpose of solving the tangled differential equations a clear exhibition is useful and helpful.
 Combining above equations together, the masses and mixings involving electron-type and sterile neutrino 
 are in principle  calculable, which further provide  input to solve the mixings involving muon and tau neutrinos, 
\begin{equation}
\begin{split}
& \frac{\dd}{\dd a}|V_{\mu1}|^2 = 2  {\text{Re}} \left[ V_{\mu1}^* V_{e1}\Big( V_{\mu2}V_{e2}^* (\Delta \tilde{m}_{12}^2)^{-1}-V_{\mu3}V_{e3}^* (\Delta \tilde{m}_{31}^2)^{-1}-V_{\mu4}V_{e4}^* (\Delta \tilde{m}_{41}^2)^{-1}\Big)\right]\\
&\hspace{1.5cm} +2 k {\text{Re}} \left[ V_{\mu1}^* V_{s1}\Big( V_{\mu2}V_{s2}^* (\Delta \tilde{m}_{12}^2)^{-1}
-V_{\mu3}V_{s3}^* (\Delta \tilde{m}_{31}^2)^{-1}-V_{\mu4}V_{s4}^* (\Delta \tilde{m}_{41}^2)^{-1}\Big)\right],\\
& \frac{\dd}{\dd a}|V_{\mu2}|^2 = 2  {\text{Re}} \left[ V_{\mu2}^* V_{e2}\Big( V_{\mu1}V_{e1}^* (\Delta \tilde{m}_{21}^2)^{-1}-V_{\mu3}V_{e3}^* (\Delta \tilde{m}_{32}^2)^{-1}-V_{\mu4}V_{e4}^* (\Delta \tilde{m}_{42}^2)^{-1}\Big)\right]\\
&\hspace{1.5cm} +2 k {\text{Re}} \left[ V_{\mu2}^* V_{s2}\Big( V_{\mu1}V_{s1}^* (\Delta \tilde{m}_{21}^2)^{-1}
-V_{\mu3}V_{s3}^* (\Delta \tilde{m}_{32}^2)^{-1}-V_{\mu4}V_{s4}^* (\Delta \tilde{m}_{42}^2)^{-1}\Big)\right],\\
& \frac{\dd}{\dd a}|V_{\mu3}|^2 = 2  {\text{Re}} \left[ V_{\mu3}^* V_{e3}\Big( V_{\mu2}V_{e2}^* (\Delta \tilde{m}_{32}^2)^{-1}+V_{\mu1}V_{e1}^* (\Delta \tilde{m}_{31}^2)^{-1}-V_{\mu4}V_{e4}^* (\Delta \tilde{m}_{43}^2)^{-1}\Big)\right]\\
&\hspace{1.5cm} +2 k {\text{Re}} \left[ V_{\mu3}^* V_{s3}\Big( V_{\mu2}V_{s2}^* (\Delta \tilde{m}_{32}^2)^{-1}
+V_{\mu1}V_{s1}^* (\Delta \tilde{m}_{31}^2)^{-1}-V_{\mu4}V_{s4}^* (\Delta \tilde{m}_{43}^2)^{-1}\Big)\right],\\& \frac{\dd}{\dd a}|V_{\mu4}|^2 = 2  {\text{Re}} \left[ V_{\mu4}^* V_{e4}\Big( V_{\mu2}V_{e2}^* (\Delta \tilde{m}_{42}^2)^{-1}+V_{\mu3}V_{e3}^* (\Delta \tilde{m}_{43}^2)^{-1}+V_{\mu1}V_{e1}^* (\Delta \tilde{m}_{41}^2)^{-1}\Big)\right]\\
&\hspace{1.5cm} +2 k {\text{Re}} \left[ V_{\mu4}^* V_{s4}\Big( V_{\mu2}V_{s2}^* (\Delta \tilde{m}_{42}^2)^{-1}
+V_{\mu3}V_{s3}^* (\Delta \tilde{m}_{43}^2)^{-1}+V_{\mu1}V_{s1}^* (\Delta \tilde{m}_{41}^2)^{-1}\Big)\right].\\
\end{split}\label{eq:VmuSquare}
\end{equation}
The differential equations on tau flavor has been neglected here as they take the similar form as  muon.
Among all the equations in the whole equation group,  equations (\ref{eq:EffMass}) and (\ref{eq:VeSquare})
play a key role to solve all the equations. 
The sterile neutrino contributions are characterized both by matter parameter $k$
and the mixings involving the fourth row and column. 
By vanishing the sterile neutrino related mixing matrix elements, $V_{\alpha 4} (\alpha=e,\ldots, s)$ and $V_{si} (i=1,\ldots 4)$ , 
equations (\ref{eq:EffMass}) and (\ref{eq:VeSquare}) return to 
standard three-flavor case \cite{Xing:2018lob}.


\subsection{Effective Jarlskog-like invariants}

The Jarlskog invariant \cite{Jarlskog:1985ht}, defined as the imaginary part of products of mixing matrix elements, 
is an important measurement to
characterize the CP violation in fermion sector. 
It is known that there is only one independent invariant in  models with  fermion family number being three. 
Previously there have been some discussions on Jarlskog-like invariants beyond three generation of quarks \cite{Hou:2010wf}
and leptons \cite{Xing:2001bg}.  Here in this work by
incorporating a light sterile neutrino, we define
 the effective Jarlskog-like invariants 
\begin{equation}
		\mathcal{J}^{ij}_{\alpha\beta}={\rm{ Im}}(U_{\alpha i} U_{\beta j} U^*_{\alpha j} U^*_{\beta i}), \qquad
                 \tilde{\mathcal{J}}^{ij}_{\alpha\beta}={\rm{ Im}}(V_{\alpha i} V_{\beta j} V^*_{\alpha j} V^*_{\beta i})
	\label{eq:Jark}
\end{equation}
in vacuum as $\mathcal{J}$ and its counterpart  $\tilde{\mathcal{J}}$ in matter, 
in which $\alpha, \beta = e,\mu, \tau, s$ and $i, j = 1,2,3,4$. 
 From the fact $\mathcal{J}_{\alpha \alpha}^{i j}=\mathcal{J}_{\alpha \beta}^{i i}=0$ and
 \begin{equation}
 \sum_i \mathcal{J}^{ij}_{\alpha \beta}=\sum_j \mathcal{J}^{ij}_{\alpha \beta}
 =\sum_\alpha \mathcal{J}^{ij}_{\alpha \beta}=\sum_\beta \mathcal{J}^{ij}_{\alpha \beta}=0
 \label{eq:J4sum}
 \end{equation}
 due to the unitary condition of the mixing matrix, the number of independent Jarlskog-like invariants increases to $9$
 compared with $1$ in the three-flavor case, which holds for the   counterpart  in matter.


The differential equation for the Jarlskog-like invariants can be derived by using equation (\ref{eq:Vdot}) after a straightforward algebraic calculation,
\begin{eqnarray}
\frac{\dd}{\dd a}\tilde{\mathcal{J}}^{ij}_{\alpha\beta} 
&=& \sum_{k\neq i} {\rm {Im}}\left[
V^*_{\alpha j} V^*_{\beta i}V_{\alpha k}  V_{\beta j} ( V^*_{e k} V_{ei} + k V^*_{sk} V_{si})
\right] 
(\Delta \tilde{m}^2_{ik})^{-1}
\label{eq:J4}\\
&& + \sum_{k\neq i} {\rm{Im}}
\left[ V_{\alpha i} V_{\beta j} V^*_{\alpha j} V^*_{\beta k} 
( V_{ek} V^*_{ei} + k V_{sk}V^*_{si})
\right] (\Delta \tilde{m}^2_{ik})^{-1}
+ (\alpha \leftrightarrow \beta, i\leftrightarrow j).\nonumber
\end{eqnarray}
As an example, we take 
 $(\alpha\beta, ij)=(e\mu, 12)$ and show explicitly  the evolution equation of one 
of the Jarlskog-like invariants in matter,
\begin{equation}
\begin{split}
\frac{d}{da}\tilde{\mathcal{J}}_{e\mu}^{12}=&\quad\;\tilde{\mathcal{J}}_{e\mu}^{12}\left(\frac{|V_{e2}|^2}{\Delta\tilde{m}^2_{12}}+\frac{|V_{e3}|^2}{\Delta\tilde{m}^2_{13}}+\frac{|V_{e4}|^2}{\Delta\tilde{m}^2_{14}}\right)
+\frac{|V_{e1}|^2}{\Delta\tilde{m}^2_{13}}\tilde{\mathcal{J}}_{e\mu}^{32}+\frac{|V_{e1}|^2}{\Delta\tilde{m}^2_{14}}\tilde{\mathcal{J}}_{e\mu}^{42}
\\
&+\tilde{\mathcal{J}}_{e\mu}^{12}\left(\frac{|V_{e1}|^2}{\Delta\tilde{m}^2_{21}}+\frac{|V_{e3}|^2}{\Delta\tilde{m}^2_{23}}+\frac{|V_{e4}|^2}{\Delta\tilde{m}^2_{24}}\right)
+\frac{|V_{e2}|^2}{\Delta\tilde{m}^2_{23}}\tilde{\mathcal{J}}_{e\mu}^{13}+\frac{|V_{e2}|^2}{\Delta\tilde{m}^2_{24}}\tilde{\mathcal{J}}_{e\mu}^{14}
\\
&+k\left(\frac{|V_{e2}|^2\tilde{\mathcal{J}}_{\mu s}^{21}}{\Delta\tilde{m}^2_{12}}+\frac{{\textrm{Im}}\big{[}V_{\mu2}V_{e2}^{\ast}V_{\mu1}^{\ast}V_{e3}V_{s1}V_{s3}^{\ast}\big{]}}{\Delta\tilde{m}^2_{13}}+\frac{{\textrm{Im}}\big{[}V_{\mu2}V_{e2}^{\ast}V_{\mu1}^{\ast}V_{e4}V_{s1}V_{s4}^{\ast}\big{]}}{\Delta\tilde{m}^2_{14}}\right)\\
&+k\left(\frac{|V_{\mu2}|^2\tilde{\mathcal{J}}_{es}^{12}}{\Delta\tilde{m}^2_{12}}+\frac{{\textrm{Im}}\big{[}V_{e1}V_{\mu2}V_{e2}^{\ast}V_{\mu3}^{\ast}V_{s3}V_{s1}^{\ast}\big{]}}{\Delta\tilde{m}^2_{13}}+\frac{{\textrm{Im}}\big{[}V_{e1}V_{\mu2}V_{e2}^{\ast}V_{\mu4}^{\ast}V_{s4}V_{s1}^{\ast}\big{]}}{\Delta\tilde{m}^2_{14}}\right)\\
&+k\left(\frac{|V_{e1}|^2\tilde{\mathcal{J}}_{\mu s}^{21}}{\Delta\tilde{m}^2_{21}}+\frac{{\textrm{Im}}\big{[}V_{e1}V_{\mu2}V_{\mu1}^{\ast}V_{e3}^{\ast}V_{s3}V_{s2}^{\ast}\big{]}}{\Delta\tilde{m}^2_{23}}+\frac{{\textrm{Im}}\big{[}V_{e1}V_{\mu2}V_{\mu1}^{\ast}V_{e4}^{\ast}V_{s4}V_{s2}^{\ast}\big{]}}{\Delta\tilde{m}^2_{24}}\right)\\
&+k\left(\frac{|V_{\mu1}|^2\tilde{\mathcal{J}}_{es}^{12}}{\Delta\tilde{m}^2_{21}}+\frac{{\textrm{Im}}\big{[}V_{e1}V_{e2}^{\ast}V_{\mu1}^{\ast}V_{\mu3}V_{s2}V_{s3}^{\ast}\big{]}}{\Delta\tilde{m}^2_{23}}+\frac{{\textrm{Im}}\big{[}V_{e1}V_{e2}^{\ast}V_{\mu1}^{\ast}V_{\mu4}V_{s2}V_{s4}^{\ast}\big{]}}{\Delta\tilde{m}^2_{24}}\right),
\end{split}\label{eq:Jeu12}
\end{equation}
which recovers the standard three-active-flavor case in the limit of $\tilde{m}_4\to \infty$ and $k\to 0$.
More Jarlskog-like invariants enter equation (\ref{eq:Jeu12}) compared with the general form, equation (\ref{eq:J4}), 
since the special 
roles of electron and sterile neutrino play.  
In the following numerical analysis, we will present evolutions of all the $9$ Jarlskog-like invariants completely. 

In addition, 
it can be verified that the relation
\begin{equation}
\sum_{i=1}^4\frac{\dd}{\dd a}\ln |V_{ei}|^2 + 2 \sum_{j>k}\frac{\dd }{\dd a} \ln \Delta \tilde{m}^2_{jk}=0
\end{equation}
holds in the limit of $k=0$, corresponding to equation (16) of \cite{Xing:2018lob}.
On the other hand, the Naumov-like \cite{Naumov:1991ju} 
and Toshev-like \cite{Toshev:1991ku} relation in RGE approach will be explored in
a separate work \cite{Naumov4} due to its complication. 

\subsection{General formulae for oscillation probabilities}
The mixing matrix and mass square differences calculated above 
are not physical observables themselves.  In fact,  they are encoded in neutrino oscillations.
By replacing the  mixing and 
mass  in the vacuum with corresponding ones in matter,
the general form of neutrino oscillation probability  can be obtain
\begin{equation}
P(\nu_\alpha\to\nu_\beta)= \sum_i |V_{\alpha i}|^2|V_{\beta i}|^2
+ 2 \sum_{i<j}\left[ {\text{Re}}(V_{\alpha i }V_{\beta j} V_{\alpha j}^* V_{\beta i}^*)\cos \tilde{\Delta}_{ij} 
-{\textrm{Im}}(V_{\alpha i}V_{\beta j} V_{\alpha j}^* V_{\beta i}^*)
\sin \tilde{\Delta}_{ij} 
\right]\label{eq:P}
\end{equation}
with $\tilde{\Delta}_{ij}\equiv \frac{\Delta\tilde{m}^2_{ij} L}{2E}$. 
%
We will see equation (\ref{eq:P}) is the most generic expression to incorporate all 
situations of the neutrino oscillation supported by the following facts:
\begin{itemize}

\item Making use of the replacement rule $V_{ij}\leftrightarrow V^*_{ij}$, the oscillations among antineutrinos 
can  also be calculated by 
equation (\ref{eq:P}).


\item Apparently oscillations occurring in both vacuum and matter are applicable, which depends on the choice of mixing and 
mass parameters, $(U, m_i)$ or $(V, \tilde{m}_i)$.

\item It can be utilized to describe oscillations among both in  three pure active flavor case and
three active together with one sterile case.
The former one can be easily trended to by vanishing $V_{si}$ ($i=1,\ldots,4)$ and $V_{\alpha 4}$ ($\alpha= e, \ldots, s)$.

\end{itemize}

So far analytical tools for further exploring neutrino phenomenology have been prepared.
We will examine this method in practical numerical study in the following part. 

\section{Applications and numerical results}
\label{sec:num}
To establish the connection between RGE and neutrino matter effects is more than a concept formally.
In practical qualitative analysis, the RGE approach is also valuable and we will realize it here.
In this section, we first numerically analyze the  evolution of effective mixing matrix explicitly . Then 
we  utilize the general oscillation eq. (\ref{eq:P}) together with RGE approach to effective mixing and mass in Earth and solar matters
to calculate corresponding observables and the numerical results are presented, respectively.

\subsection{The evolution of mixing matrix elements}

The propagation of neutrino  in various  mediums depends on effective mixing matrix 
in different materials, hence it is necessary to perform an explicit analysis of detailed mixing
matrix elements firstly. In constant density matters, the features of matter effects
are  controlled by the matter parameter $k$. For example, 
 it is known that  $k=-0.5$ depicts  earth matter. 
By adopting the input parameters in Appendix \ref{app:input},  we show in Fig.\ref{fig:Vsquare}  the evolution of effective mixing matrix elements
with particular $k$  in normal hierarchy (NH)  case  with respect to 
energy scale $a/\Delta m^2_{21}$, where
 neutrinos and antineutrinos are characterized as solid red and
dashed blue curves, respectively.

\begin{figure}[t!]
	\centering 
	$
	\begin{array}{c}
		(a): k=0 \\
	\includegraphics[width=0.88\linewidth]{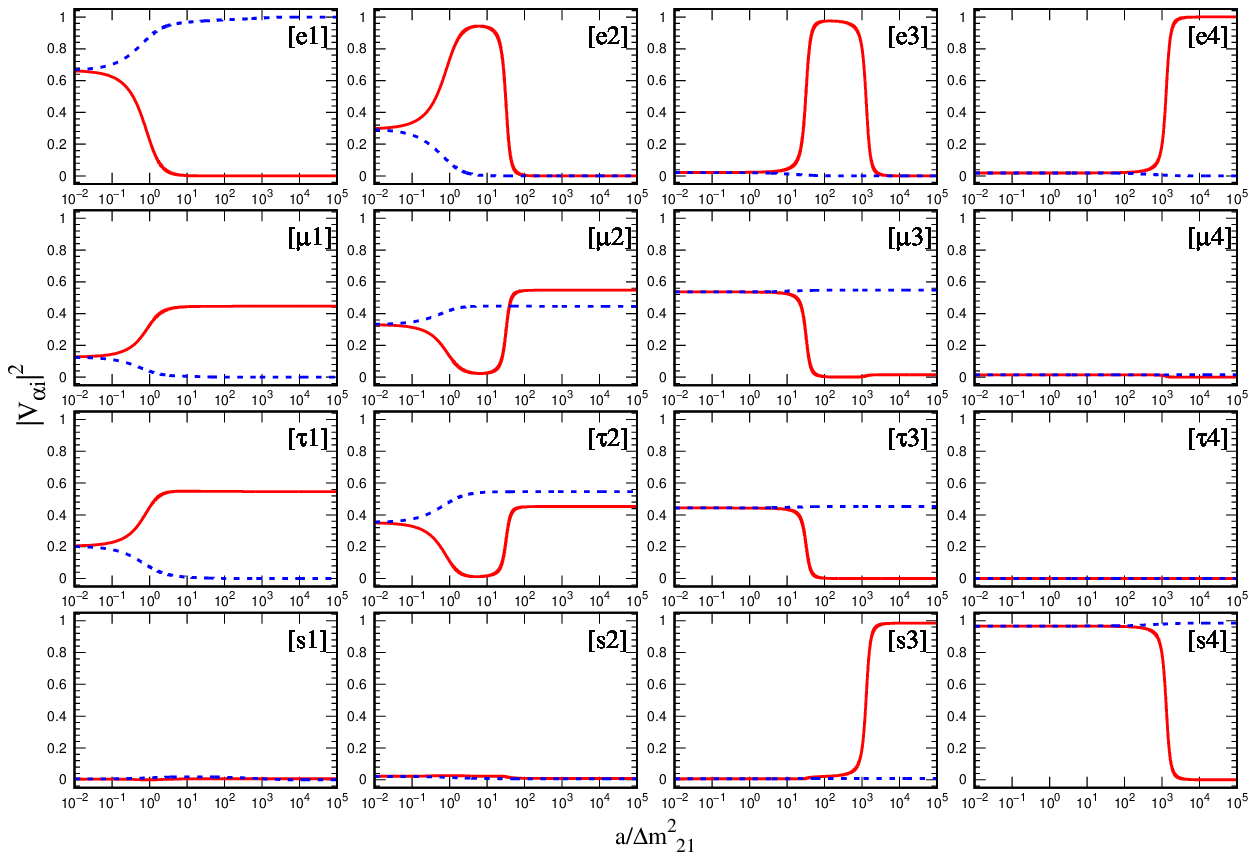}\\
	    (b): k= - 0.5\\
         \includegraphics[width=0.88\linewidth]{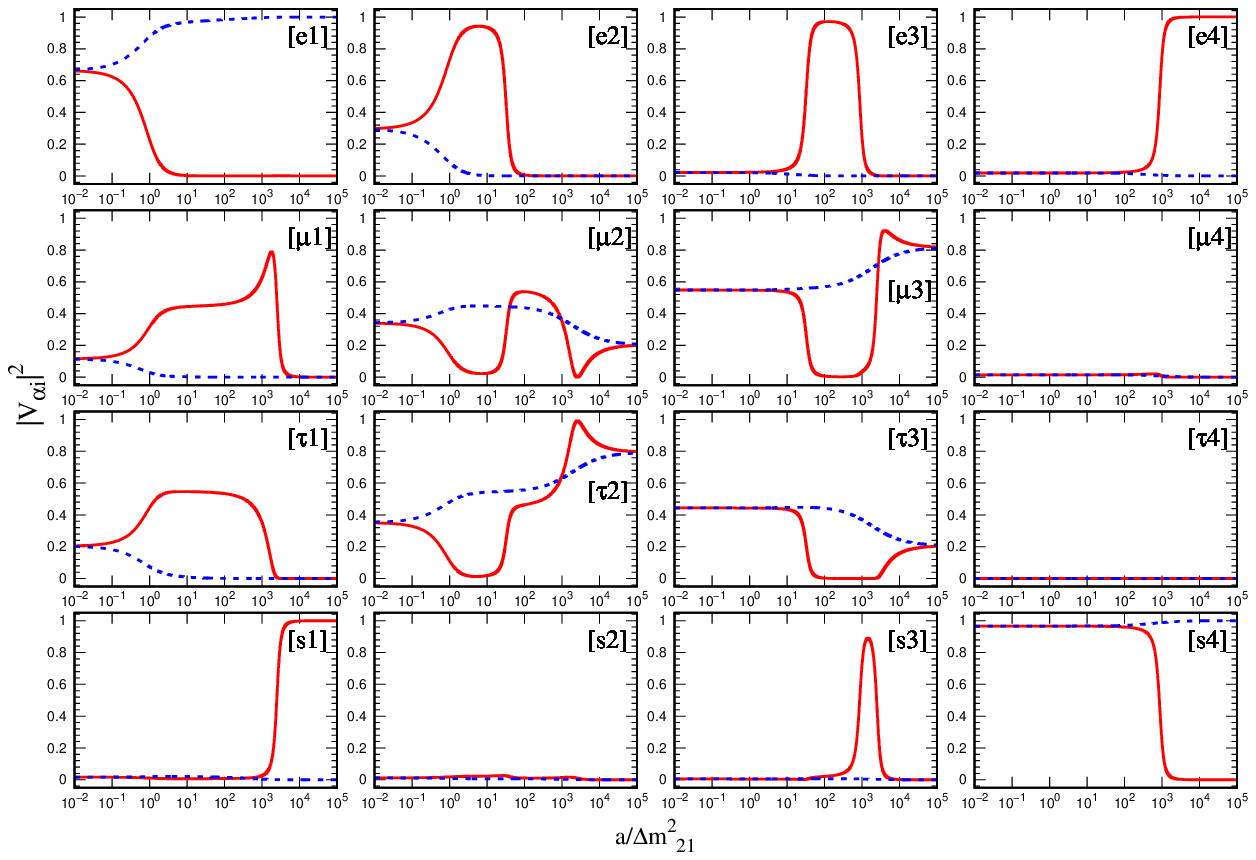}\\
        \end{array}
       $
    \caption{The evolution of $|V_{\alpha i}|^2$ (for $\alpha=e,\mu,\tau, s$ and $i=1,2,3,4$) in NH case 
    with respect to 
    energy scale $a/\Delta m^2_{21}$, in which: $ k=0$  in $(a)$   and $ k=-0.5$ in $(b)$.
    The red  solid and blue dashed curves correspond to the results of 
     neutrino and antineutrino oscillations, respectively.}  
     \label{fig:Vsquare}
\end{figure}

The situation for $k=0$ is special. 
For an illustration, we take the first row of  Fig.\ref{fig:Vsquare}(a) as an example to 
interpret the evolutions of $|V_{\alpha i}|^2$ according to equation (\ref{eq:VeSquare}). 
The negative (positive) sign of the right hand side 
 of $|V_{e1}|^2$ ($|V_{e4}|^2$) evolution equation  indicates 
 a monotonic increasing (decreasing) behavior 
 for  $|V_{e1}|^2$ ($|V_{e4}|^2$). Due to a competition between positive and 
 negative parts in the RGE equations, combining their particular initial values, the maximums
 appear  for the evolution of $|V_{e2}|^2$
 and $|V_{e3}|^2$.
In the large $\Delta \tilde{m}_{41}^2$ limit, the sterile neutrino
contribution is decoupled and the evolutions  return to  three  pure active flavor case.
\footnote{The standard three-flavor limit can also be obtained by setting $V_{\alpha 4}=0$ and 
$V_{s i}=0$ as aforementioned. Here we understand $3\nu$ picture in this way is for the purpose of examining corresponding
results in  \cite{Xing:2018lob}.}
Comparing the left three columns of the rows except the last one (the mixings
among three active flavors) in
Fig.\ref{fig:Vsquare}(a) with  the plot Fig.1 of \cite{Xing:2018lob}, one can observe the behaviors of evolution
curves are consistent with each other. 
Both the evolution behaviors and initial values  of $|V_{\mu i}|^2$, 
standing for vacuum matrix elements $|U_{\mu i}|^2$, are qualitatively 
identical to  $|V_{\tau i}|^2$  ($i=1,2,3$).
The  approximate $\mu-\tau$ symmetry is kept and 
 well understood as the flavors are indistinguishable for muon and tau
for the universal neutral-current interactions of the two flavors in ordinary matter.
Mathematically, this can also be explained
by setting zero values to $k$ and $(\Delta \tilde{m}^2_{41})^{-1}$
in the first three equations of  (\ref{eq:VeSquare}) and  (\ref{eq:VmuSquare}).
Among all the six active-sterile mixing matrix elements for neutrino case (red solid curves), 
only two of them 
$|V_{e4}|^2$ and $|V_{s3}|^2$  manifest themselves as unity
 when energy scale $a$ reaches around $m_4^2$ 
 (or $a/\Delta m^2_{21}\sim \Delta {m}^2_{41}/\Delta m^2_{21}\sim1000$ ).
 In contrast, there is no
 mixing among active and sterile flavors for antineutrinos.  For the pure
 sterile neutrino, $|V_{s4}|^2$ keeps unity and drops to small value near to zero when $a$
 approaches $\Delta {m}_{41}^2\sim 10^3 \Delta m^2_{21}$, which guarantees 
 the unitary conditions $\sum\limits_{i=1}^4 |V_{\alpha i}|^2 =1$ for a particular flavor $\alpha$ 
and $\sum\limits_{\alpha=e}^s |V_{\alpha i}|^2 =1$ for a particular  mass index $i$.

The case of $k=-0.5$, corresponding to the terrestrial matter, 
contains the contribution from non-vanishing $k$ terms, which reflects complete
sterile neutrino contributions  in constant density matter. In addition to
the well respected unitary conditions, some obvious changes occur  in Fig.\ref{fig:Vsquare}(b):
\begin{itemize}
\item The effective $\mu-\tau$ symmetry is partially breaking  for $i=1, 2, 3$ in neutrino case and  $i=3,4$ for antineutrino case.
\item While keeping the active-sterile mixings unchanged from the $k=0$ scenario of antineutrino case, two of the six mixing matrix elements 
of neutrino case ($|V_{s1}|^2$ and $|V_{s3}|^2$) change their behaviors.  

\item The value of $|V_{s4}|^2$ is stably kept around one since the mixing among sterile neutrino and other ones is close to
zero for the antineutrino case.  The evolution for neutrino case differs as a sudden drop from unity to zero occurs around $10^3$, which is consistent with the behaviors
of mixing among sterile neutrino and the active ones.
\end{itemize}

Similarly, the inverse mass hierarchy (IH) case 
as well as the evolution of effective masses, described in Eq. (\ref{eq:EffMass}), can also be analyzed and 
we have neglected the corresponding plots. 
Now with the  effective mixing matrix and mass square prepared, we may apply these results to further analyze neutrino propagation
in typic mediums. 

\subsection{The evolution of Jarlskog-like invariants}

\begin{figure}[t]
	\centering 
	\includegraphics[width=\linewidth]{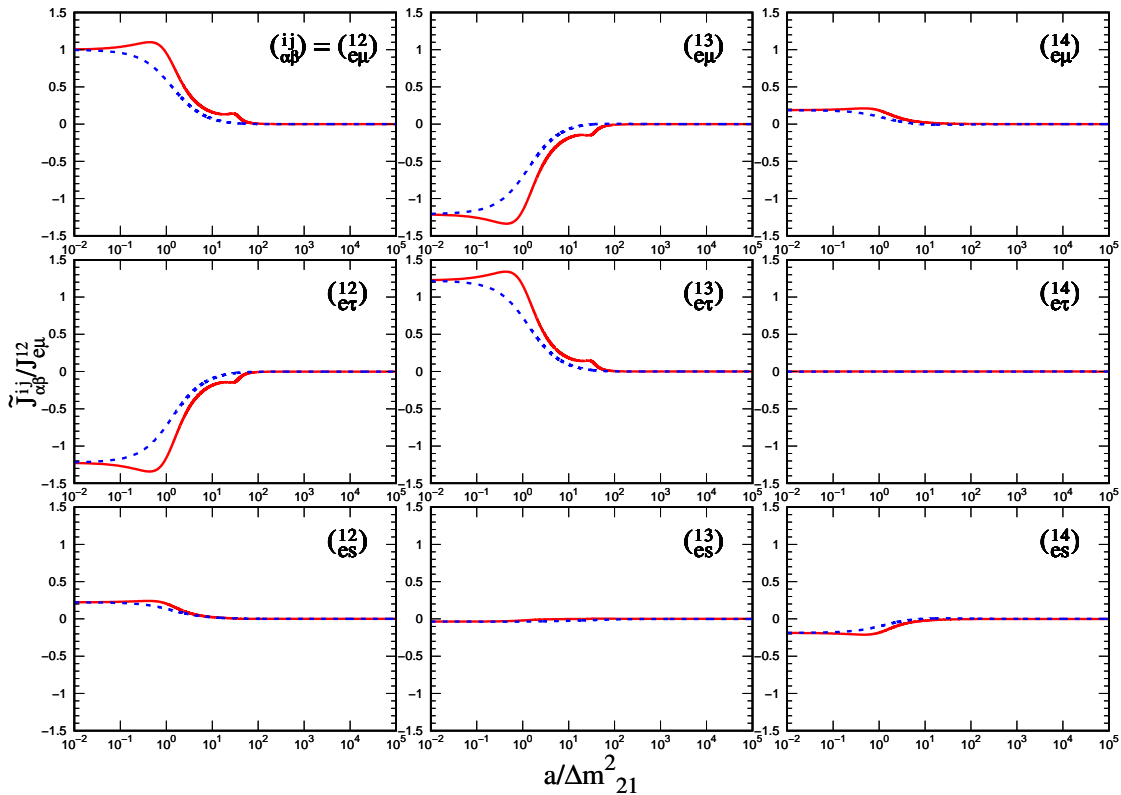}
    \caption{The evolution of the Jarlskog-like invariants $\tilde{\mathcal{J}}^{ij}_{\alpha\beta}$ in NH case, normalized by $\mathcal{J}^{12}_{e\mu}$, with respect to $a/\Delta m^2_{21}$.  The red  solid and blue dashed curves correspond to the results of 
     neutrino and antineutrino oscillations, respectively.}  
     \label{fig:J}
\end{figure}

For the nine Jarlskog-like invariants, the choice of the basis is not unique. Here we fix one of the flavor and mass indices as 
electron and  $1$ and let the remaining indices run over all the paramete,sr space.  
In Fig. \ref{fig:J}, within NH framework by setting $k = 0$ we show the evolution behaviors of these nine invariants normalized to
the vacuum value of one invariant $\mathcal{J}_{e\mu}^{12}$ with respect to $a/\Delta m^2_{21}$. 

Typically the  behavior of the Jarlskog-like invariant  in the first plot of Fig. \ref{fig:J} can be interpreted 
from equation (\ref{eq:J4}). 
Among all the terms in the right hand side,  terms containing  $\Delta\tilde{m}^2_{21}$ dominate
the evolution trend. During the initial stage,  the increase of $\tilde{\mathcal{J}}_{e\mu}^{12}$
is determined by positivity of $(|V_{e1}|^2-|V_{e2}|^2)\Delta \tilde{m}^2_{21}$. 
At around $a/\Delta m^2_{21}\sim 1$,  the first resonance is approached for the sign flip of
$(|V_{e1}|^2-|V_{e2}|^2)$, see the first two plots in Fig. \ref{fig:Vsquare}(a).
And then a second peak appears at around $a/\Delta m^2_{21}\sim 10$, 
due to the change of the relative size between $|V_{e2}|^2$ and $|V_{e3}|^2$, which 
is consistent with the pure three-flavor situation.
Analogically the other eight invariants can be analyzed and here we exhibit
their evolutions as the remaining parts of Fig. \ref{fig:J}, in which
 the sum rules in equation (\ref{eq:J4sum}) are respected in each row or column.

\subsection{The matter effects in Earth}

Generally speaking, terrestrial matter effects should  be taken into account in all the experiments taken place in Earth.
Practically, there is few difference in the propagation between vacuum and matter in short-baseline (SBL) and medium-baselined (MBL) experiments.
These experiments are usually built around sites nearby  nuclear power plants with electron-type antineutrino as their source, given
\begin{equation}
P(\bar{\nu}_e\to \bar{\nu}_e) = \sum_i |V_{ei}|^4+2 \sum_{i<j} |V_{ei}|^2 |V_{ej}|^2 \cos \tilde{\Delta}_{ij}.
\label{eq:SBL}
\end{equation}
Since the matter effects are not significant, the  mixing matrix and mass difference in above Eq. (\ref{eq:SBL})
can be safely  replaced by the corresponding ones in vacuum.

To further study neutrino mass ordering and measure CP-violating phase in the mixing matrix, taking the neutrino produced from accelerators with 
high energy as the experiment source is a good choice. The more energetic neutrino is supposed to propagate in a longer distance and hence
 longer baselines are required  as one of the experimental conditions. 
 In the LBL experiments, the matter effects
are no longer negligible.
 The baselines for 
two ongoing LBL experiments,
NOvA (NuMI Off-axis electron Neutrino Appearance) and T2K (Tokai to Kamioka),  are  given as
  $L=810\,{\text{km}}$ and $L=295\,{\text{km}}$, while 
 the baseline is designed to be $L=1300\,{\text{km}}$ for
the near-future Fermilab experiment DUNE (The Deep Underground Neutrino Experiment). 
In these LBL experiments with accelerator neutrino source, both  flavor changing and conserving modes  can be observed 
and more physical goals, such as the  measurement of leptonic CP-violating phase,  are supposed to be achieved.
In presence of a light sterile neutrino, the number of CP-violating phases enhances to $3$
 and hence the reliance on phases is more complicated. 
To make use of previous numerical results  
of mixing matrix and mass square difference
in  RGE approach,   the general expression for neutrino oscillation 
is preferred, giving
\begin{equation}
P({\nu}_\mu\to {\nu}_e) = 
\sum_i |V_{\mu i}|^2|V_{e i}|^2
+ 2 \sum_{i<j}\left[ {\text{Re}}(V_{\mu i }V_{e j} V_{\mu j}^* V_{e i}^*)\cos \tilde{\Delta}_{ij} 
-{\textrm{Im}}(V_{\mu i}V_{e j} V_{\mu j}^* V_{e i}^*)
\sin \tilde{\Delta}_{ij} 
\right].
\label{eq:LBL}
\end{equation}
For the three LBL experiments, though with different baselines, they share the common mixing matrix since 
the neutrinos received by detectors all travel through the crust.  
In the following analysis,  we adopt the appearance mode  as an illustration.

\begin{figure}[t!]
	\centering  
	$
	\begin{array}{cc}
	\includegraphics[width=0.5\linewidth]{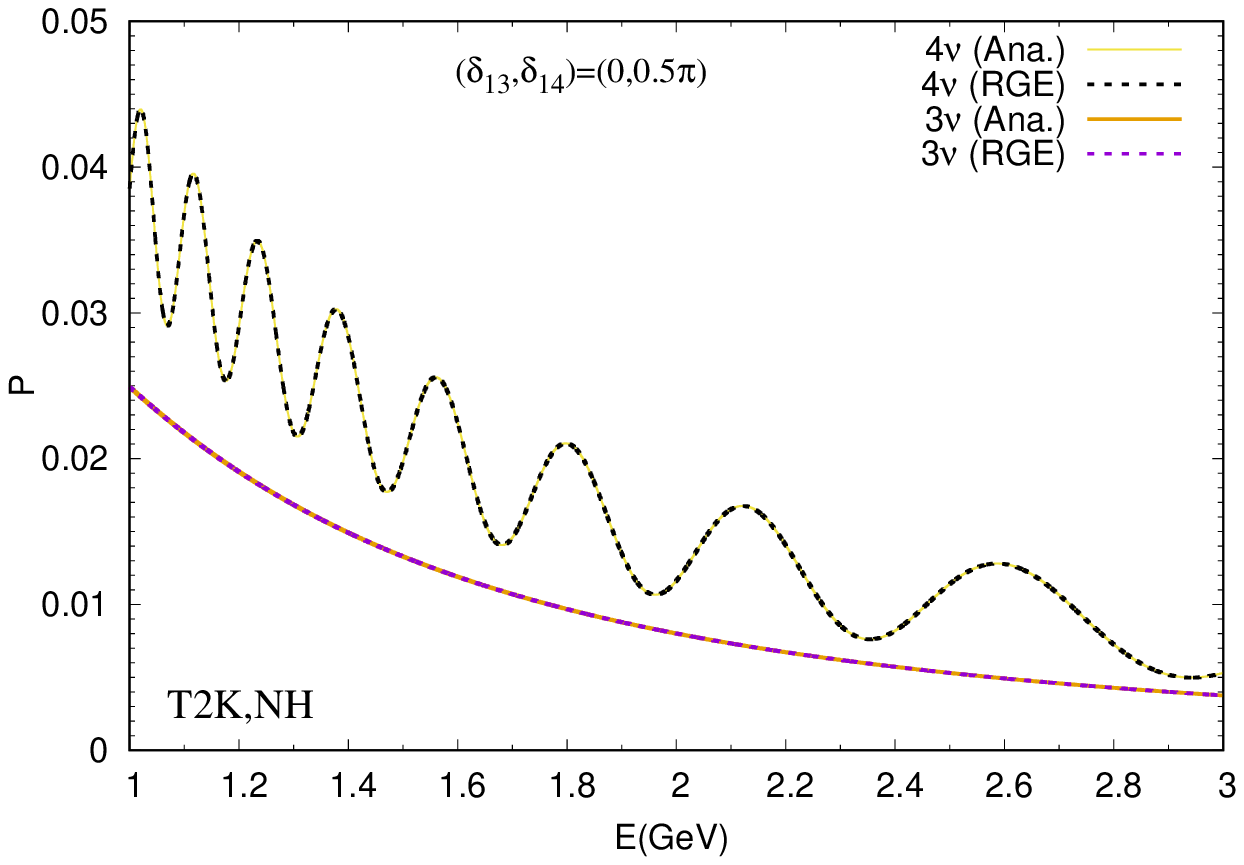} &
	\includegraphics[width=0.5\linewidth]{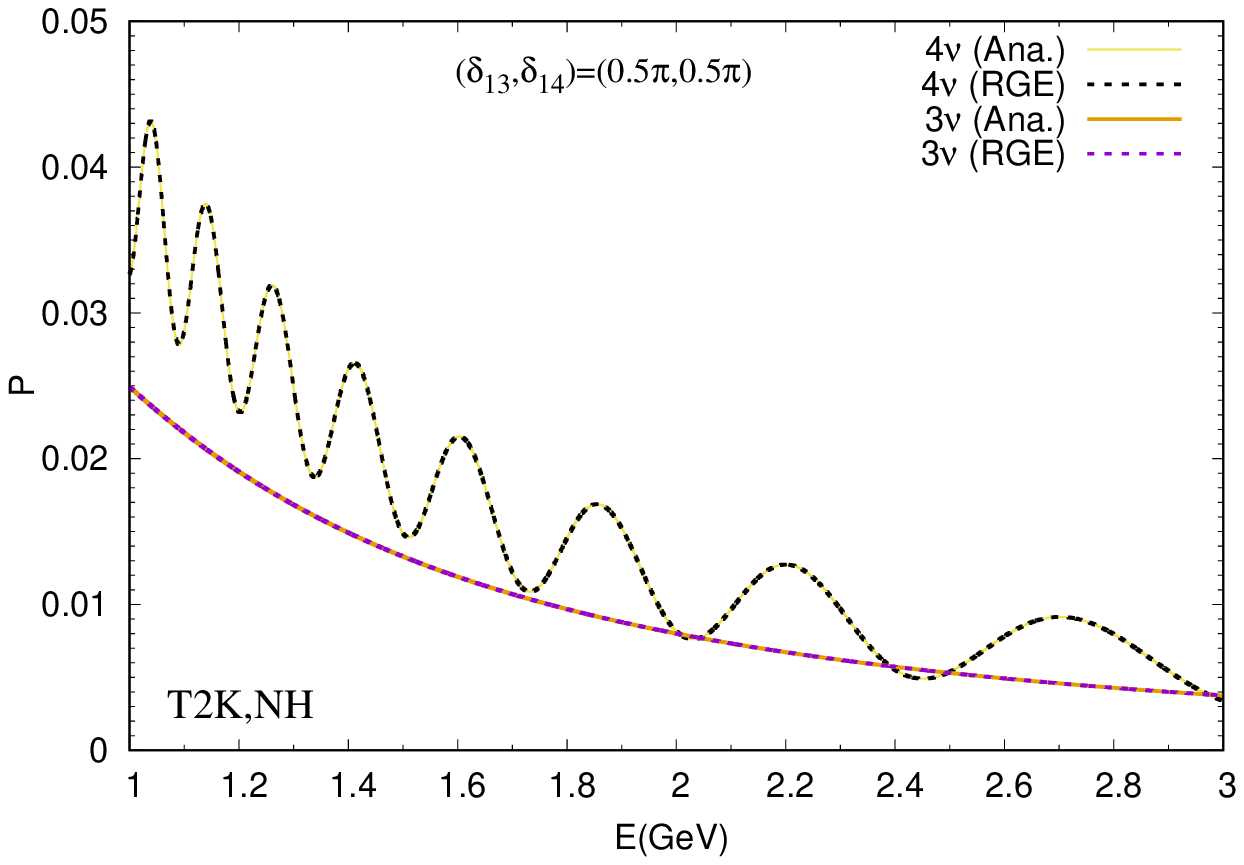} \\
	(1) & (2) \\
	\includegraphics[width=0.5\linewidth]{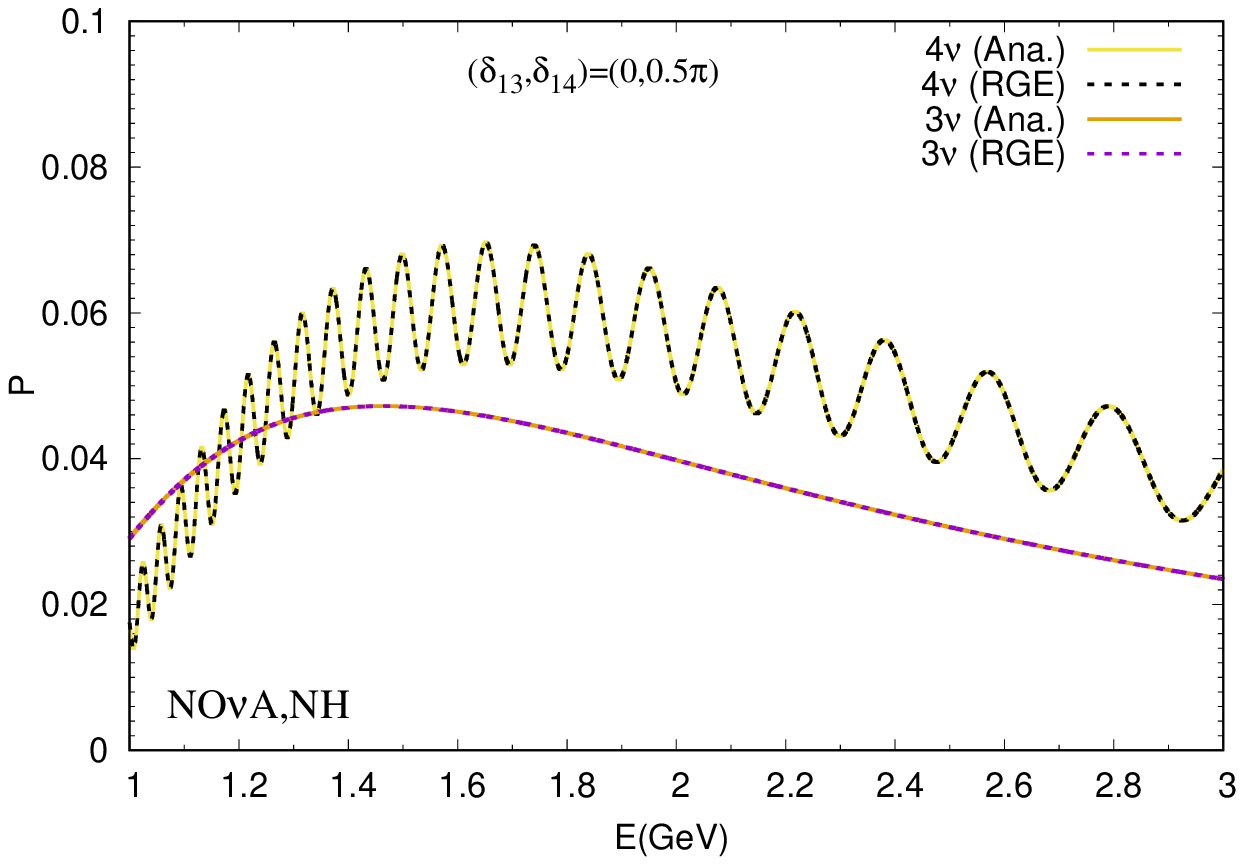} &
	\includegraphics[width=0.5\linewidth]{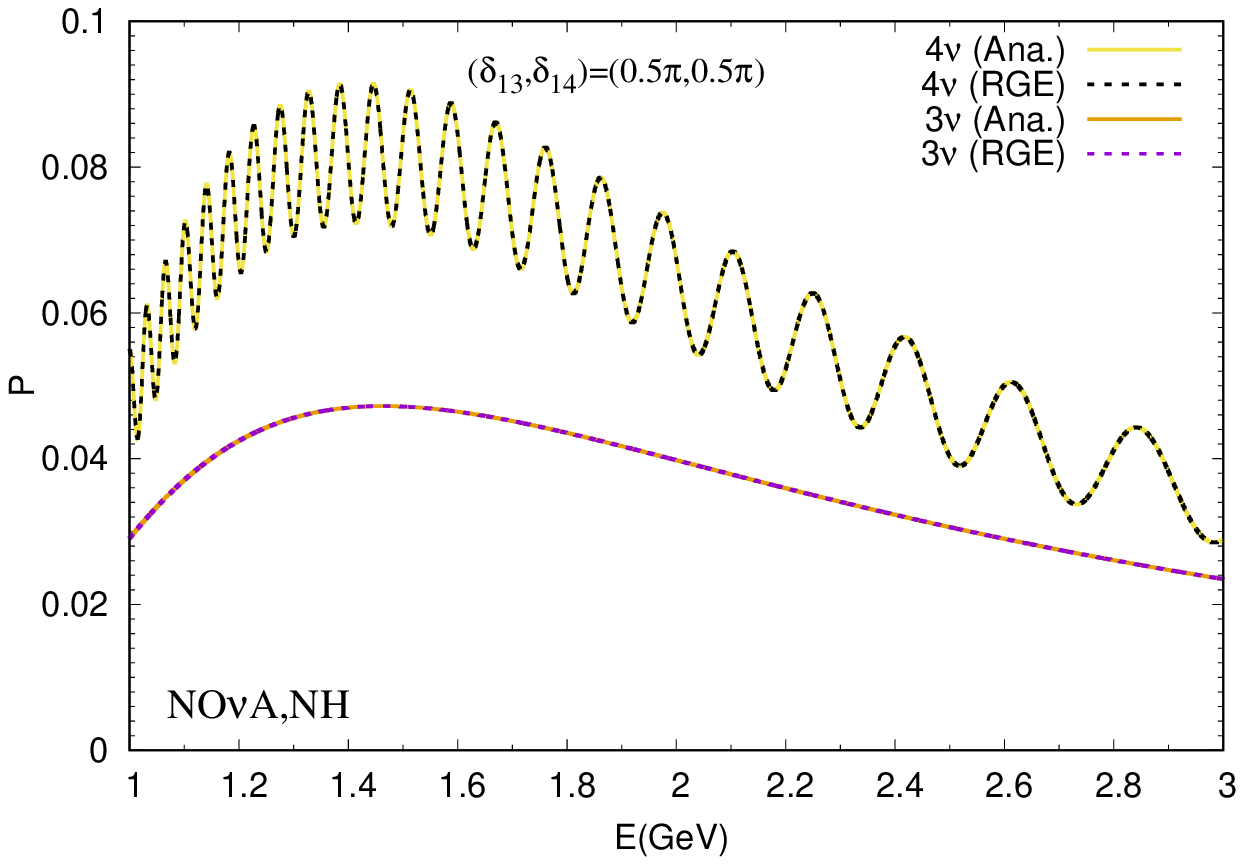} \\
	(3) & (4)\\
	\includegraphics[width=0.5\linewidth]{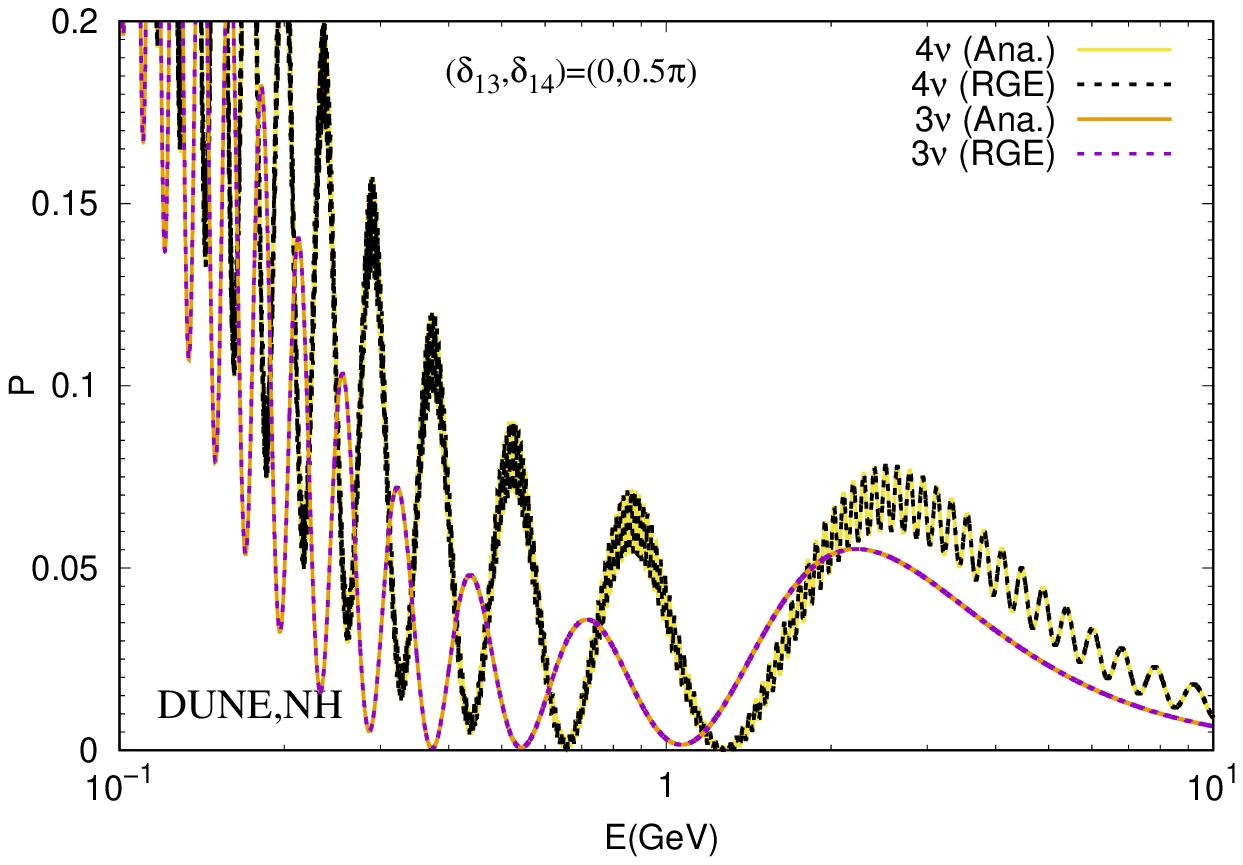} &
	\includegraphics[width=0.5\linewidth]{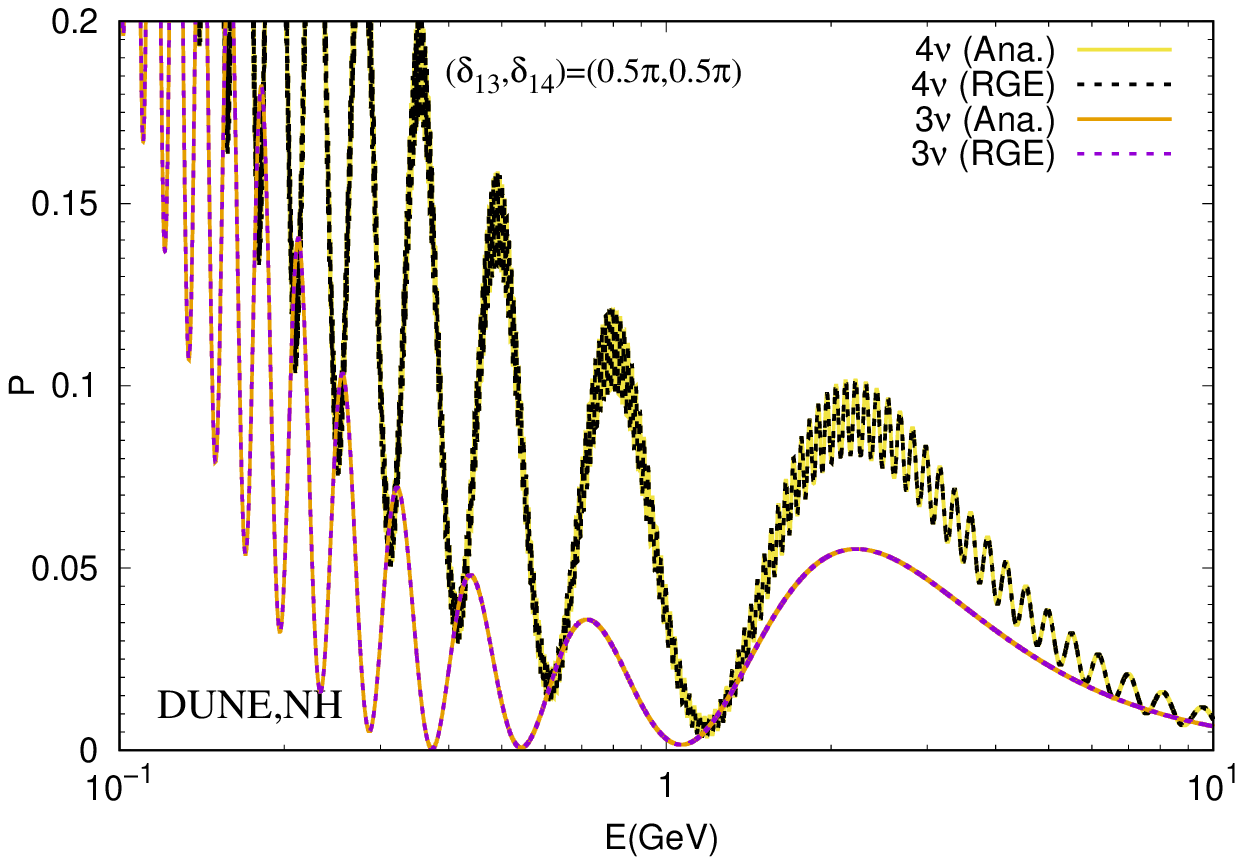} \\
	(5) & (6)
	\end{array}
	$
    \caption{The oscillation probabilities of appearance mode ${\nu}_\mu\to {\nu}_e$ in LBL accelerator neutrino experiments T2K, NOvA and DUNE for NH case with  $(\delta_{13}, \delta_{14})=(0,\frac{\pi}{2}),(\frac{\pi}{2},\frac{\pi}{2})$. }  
    \label{fig:oscillation}
\end{figure}

In the numerical calculations, the vacuum mixing (PMNS) matrix as well as particular mass-square differences are treated as the input.
We adopt the values in
\cite{Zyla:2020zbs} for three-flavor situation and a global fit result \cite{Gariazzo:2017fdh} for the four-flavor case, 
summarized in Appendix \ref{app:input}.
To produce the curve $P(E)$, the $E$ dependence of $|V_{\alpha i}|^2, V_{\alpha i}$ and  $\tilde{\Delta}_{ij}$  are required.
These quantities are calculable by integrating related differential equations derived in Section \ref{subsec:eq}.

In addition to plotting the oscillation curves of  the three LBL experiments in RGE approach, we also make use of the  exact analytical  formulae derived in \cite{Li:2018ezt}
to  guarantee the effectiveness of this quantitative analysis based on RGE, shown in Fig. \ref{fig:oscillation}.
We  find that:

\begin{itemize}

\item All the curves plotted in two different ways coincide perfectly, indicating that the RGE approach works
well in numerical study.

\item The formulae developed in this work, including sterile neutrino contribution,  can be utilized to describe four different situations
($3\nu$ in vacuum and matter, $4\nu$ in vacuum and matter) by opting for proper parameters.

\item  The matter effects are manifest for all the three LBL experiments. 
Although the orders of magnitude are of percentage level,
with sterile neutrino the values of  probability have a global enhancement.
Taking NOvA with $(\delta_{13},\delta_{14})=(0.5\pi, 0.5\pi)$ as an example (see Fig. \ref{fig:oscillation}(4)), 
the enhancement can 
reach as high as $100\%$ when the energy of neutrino beam is around $1.5  {\textrm{GeV}}$. 
Obviously the future precise measurement from LBL accelerator neutrino experiments will provide complementary information
of light sterile neutrino.
\end{itemize}

Now we have verified that the RGE approach is workable in quantitative study of neutrino oscillation in single constant
density matter. 
We will explore the wider application with more complicated structure in the following part.

\subsection{The  day-night asymmetry of solar neutrino}

Recently there have been some progresses in the study of solar neutrino.
Neutrinos
produced in the CNO fusion cycle have been observed
\cite{BOREXINO:2020aww}, and the potential to observe ${}^8$B
solar neutrino has been discussed
\cite{JUNO:2020hqc}.
In this part, we continue to study solar neutrino in the RGE approach as an extension
of pure terrestrial matter effects.

Before being observed on the surface of earth during  daytime, the electron-type neutrinos produced in solar core have travelled 
through solar matter, the vacuum between sun and earth, giving
the $\nu_e$ survival probabilities in the adiabatic approximation
\begin{equation}
P_D=\sum_{i=1}^4 P_{ei}^S P_{ie}^0 ,
\label{eq:day}
\end{equation}
where $P_{ei}^S$ denotes the $\nu_i$ component of $\nu_e$ produced in solar core and its
 propagation to earth surface
while $P_{ie}^0$
stands for the probabilities of electron-type neutrino contained in the $i$-th mass eigenstate.
The equation (\ref{eq:day}) is an extension of three-flavor framework \cite{Akhmedov:2004rq}, including
the contribution from
a light sterile neutrino contribution.  
Before arriving at the detectors on the other side of the incident point, solar neutrinos  pass some distance inside the earth at night 
and the extra earth matter effects should be taken into account.
The nighttime $\nu_e$ survival probability is then
\begin{equation}
P_N=\sum_{i=1}^4 P_{ei}^S P_{ie}^E , 
\label{eq:night}
\end{equation}
where $P_{ie}^E$ denotes the probability of $\nu_i$ propagation in earth matter and finally captured 
by detector in $\nu_e$ state.
 The day-night asymmetry  stands for the probability difference between the day and night time, giving
 \begin{equation}
 \Delta P=P_D-P_N=\sum_{i=1}^4 P_{ei}^S  (P_{ie}^0-P_{ie}^E).
 \end{equation}
 
 The calculation of $\Delta P$ requires the information of $P_{ei}^S, P_{ie}^0$ and $P_{ie}^E$.
 In general, we have 
\begin{equation}
P_{ie}^0=|U_{ei}|^2, \quad
P_{ei}^S=|V^S_{ei}|^2, \quad
P_{ei}^E=|V^E_{ei}|^2
\end{equation}
with vacuum PMNS matrix $U_{e i}$,  solar matter effective mixing matrix $V^S_{ei}$
and earth matter effective mixing matrix $V^E_{ei}$.
In the practical calculation, 
after producing in solar core
\footnote{In solar core, taking $N_e=102 N_A/{\textrm{cm}}^3$ for electron density \cite{Bahcall:2000nu} and
 $N_n=0.1639 N_e$ for neutron density \cite{Asplund:2009fu}, we obtain the solar core matter parameter $k=-0.5\times 0.1639
 \approx -0.08$. },
we adopt adiabatic approximation to describe neutrino propagation 
through other part of the sun and the vacuum between sun and the Earth. 
For the Earth model, we take the slab approximation 
to describe the terrestrial matter\cite{Dziewonski:1981xy}. Explicitly, the effective mixing matrix in Earth is in the form of
\begin{equation}
V^E = \left(\prod_{i=1}^N V_i e^{-i H_i L_i}V_i^\dagger\right)
U,
\label{eq:Vearth}
\end{equation}
where $N$ is the number of slabs,
$H_i$
is the effective Hamiltonian after diagonalization in $i$-th layer with corresponding neutrino traveling 
distance $L_i$ and effective mixing $V_i$.  Here 
in 8-Layer Earth Model \cite{Dziewonski:1981xy}, including the structure of Inner Core, Outer Core, 
Mantel and Crust,  the value of $N$ depends on the zenith angle.

\begin{figure}[t!]
	\centering  
	$
	\begin{array}{cc}
	\includegraphics[width=0.5\linewidth]{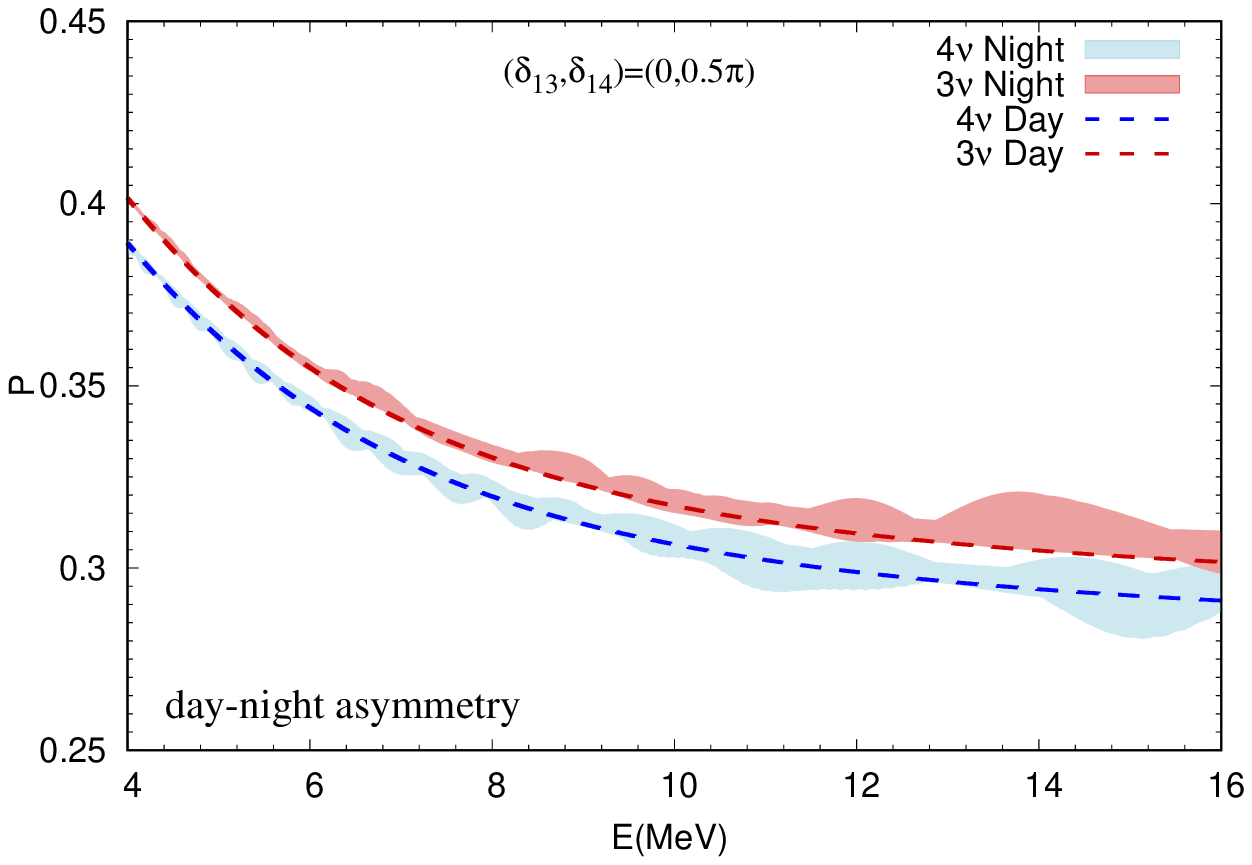} &
	\includegraphics[width=0.5\linewidth]{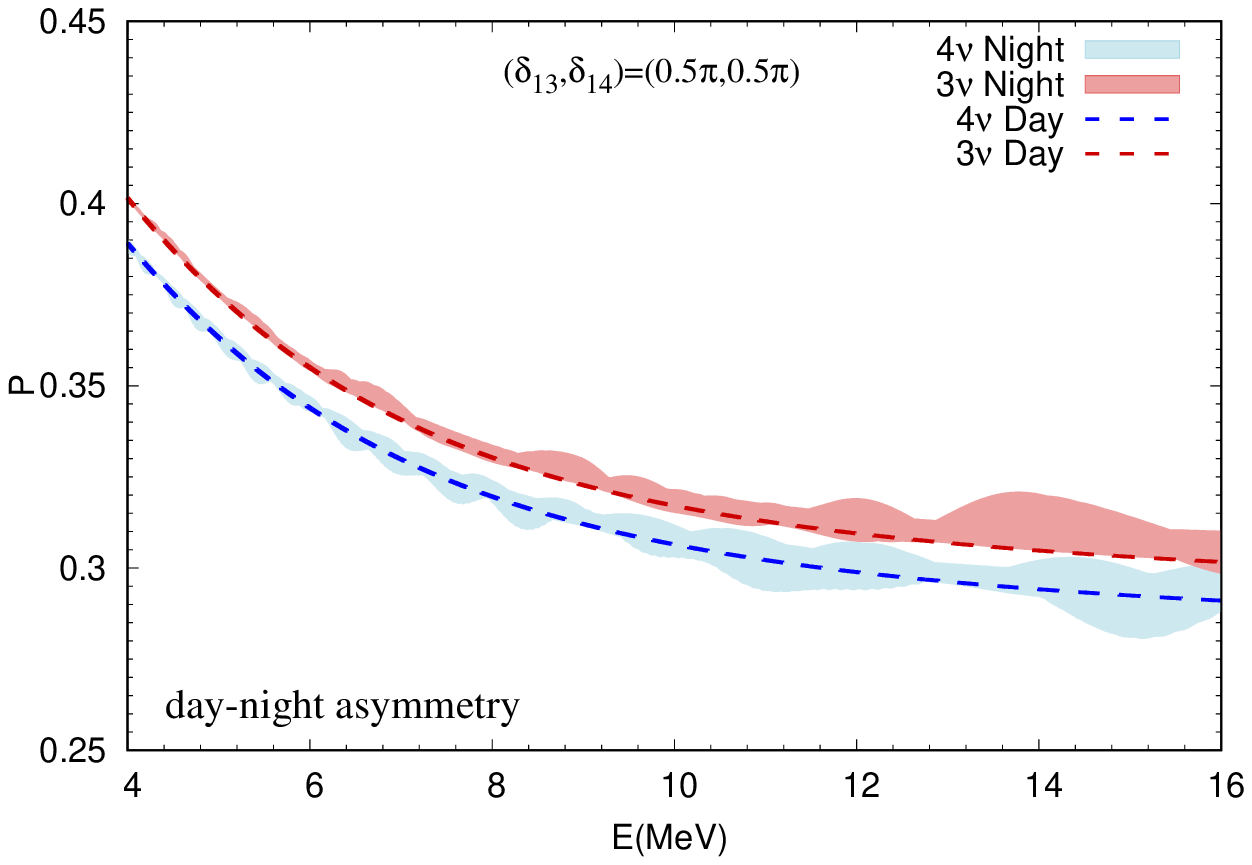} \\
	(1) & (2)
	\end{array}
	$
    \caption{Electron-neutrino survival probabilities from solar neutrino in the three-flavor (red)  and four-flavor (blue) pictures, in which
    the daytime behaviors are marked as dashed curves while nighttime ones vary in their corresponding colored ranges.}  
    \label{fig:dna}
\end{figure}

Although the mixing in Earth, equation (\ref{eq:Vearth}), is complicated due to the multiple structures of
the Earth,  RGE approach provides a proper choice for realizing the calculation. Combining all the necessary mixings both in Earth and the sun, 
the electron-neutrino survival probabilities from solar neutrino in various conditions are numerically presented in Fig. \ref{fig:dna}.
Some features can be drawn:
\begin{itemize}
\item In general, the survival probability in $4\nu$ picture  decreases compared with the corresponding one in $3\nu$ picture.
This is understandable as  the sterile neutrino takes up part of the components of electron-neutrino produced in solar core, diminishing 
the survival probability. 

\item Though  input parameter dependence exists, the  difference 
 is tiny between different parameter choices in $4\nu$ framework.
 
\item The probability at nighttime, which depends on the altitude of incident beam,  vary in a range due to the terrestrial matter effects.

\item For the maxima of day-night asymmetry $\Delta P$,  it can be reached around $13.5 \textrm{MeV}$ for three-flavor case while
$15 \textrm{MeV}$ in the four-flavor framework. This could be  further discriminated by future solar neutrino experiments.

\end{itemize}

As demonstrated in above quantitative analysis, 
 the RGE approach to neutrino matter provides  a proper tool 
 in exploring objects with richer structures.

\section{Concluding remarks}
\label{sec:con}

The RGE has a wide application in many fields of modern physics.
In the pioneering work, a connection between RGEs and neutrino matter effects
in standard three-flavor picture 
has been established and a set of differential equations 
have been obtained\cite{Xing:2018lob}.  To demonstrate the validity of RGE approach 
to matter effects in quantitative study is a worthy topic. Furthermore, to explore an extension of the entire
methodology incorporating a light sterile neutrino is also valuable for the recent fast development of sterile neutrino-related  experiments.  
These tasks have been achieved in current work.

In presence of a light sterile neutrino, we have derived a completed set of differential equations on effective mixing matrix elements,
 mass-square differences and Jarlskog-like invariants in RGE approach. 
By opting for proper input parameters (mixings and mass differences), these  equations as well as their solutions
can describe both three-flavor and four-flavor neutrino oscillation.
The mixing parameters $V_{\alpha i}$ and $\tilde{m}_i^2$
depend on the matter parameter $k$ in the constant density matter.
Solving these  differential equations, we carry out numerical analysis for the evolution of $|V_{\alpha i}|^2$ and Jarlskog-like invariants.
The consistence of three-flavor case with \cite{Xing:2018lob} is verified, which further guarantees the correctness 
for further study. 
By combining three LBL neutrino experiments, we  
calculate their corresponding oscillation probabilities in NH scenario as examples of terrestrial matter effects.
The study of electron-neutrino survival probabilities and hence day-night asymmetry of solar neutrino 
observed on the Earth surface is more complicated for types of mediums involved.
Taking the adiabatic approximation for solar matter and slab approximation for terrestrial matter, we numerically calculate the 
survival probabilities at daytime and nighttime.  
The day-night asymmetry maximizes itself at around $13.5 \textrm{MeV}$ in the standard three active flavor framework and
$15\textrm{MeV}$ in four-flavor framework, which could be discriminated in future precise measurement. 
 
The RGE approach developed in this work provides a complementary way to study 
neutrino phenomenology involving matter effects. On the other hand, the correctness of its application shown in this work
helps us to have a deep understanding of RGE.

\acknowledgments
We would like to thank Prof. Z.-z. Xing for 
his encouragement that produces the current work.
Discussions with Porf. Jiajie Ling and Prof. Benda Xu are also acknowledged. 
This work is supported by NSFC  under Grant  No. U1932104 and No. 12142502,  by 
Guangdong Provincial Key Laboratory of Nuclear Science with No. 2019B121203010.


\appendix
\section{Input parameters}
\label{app:input}
Two types of input parameters, mixing matrix and masses in both three-flavor and four-flavor pictures, are summarize 
in Table \ref{tab:input}.

\begin{table}[h!]
  \begin{center}
      \caption{A summary of the involved input parameters in the numerical analysis of this work.}
      \vspace{0.2cm}
    \begin{tabular}{c c c c c c }
    \hline
    $\sin^2\theta_{13}$&$\sin^2\theta_{12}$&$\sin^2\theta_{23}$&$\Delta{m}^2_{21}$&$\Delta{m}^2_{31}$&$\delta_{13}$\\
    $0.0220$&$0.307$&$0.546$&$7.53\times10^{-5}\textrm{eV}^2$&$2.526\times10^{-3}\textrm{eV}^2$&$1.36\pi$\\
    \hline\hline
    $\sin^2\theta_{14}$&$\sin^2\theta_{24}$&$\sin^2\theta_{34}$&$\Delta{m}^2_{41}$&$\delta_{14}$&$\delta_{34}$\\
    $0.019$&$0.015$&$0$&$0.1\textrm{eV}^2$&$0.5\pi$&$0$\\
    \hline
    \end{tabular}
    \label{tab:input}
  \end{center}
\end{table}

\begin{itemize}
\item Three-flavor framework 

We adopt the standard parameterization of PMNS matrix and the inputs are taken from PDG \cite{Zyla:2020zbs}.

\item Four-flavor framework 

We follow the convention in Appendix A of \cite{Li:2018ezt}
to parameterize the
four-dimensional mixing matrix 
with three leptonic CP-violating phases $(\delta_{13}, \delta_{14}, \delta_{34})$.
A global fit results \cite{Gariazzo:2017fdh} are taken as input here, 
in which $\delta_{13}$ is kept as a free parameter.
\end{itemize}

\bibliographystyle{JHEP}
\bibliography{reference}

\end{document}